\documentclass[prb,aps,final,twocolumn,floatfix]{revtex4}
\input epsf
\input pstricks
\begin{document}
\def\etr{\varepsilon_{\rm tr}}
\def\beq{\begin{equation}}
\def\eeq{\end{equation}}
\title{Statistical Evaporation of Rotating Clusters}
\author{F. Calvo}
\affiliation{Laboratoire de Physique Quantique, IRSAMC, Universit\'e Paul
Sabatier, 118 Route de Narbonne, F31062 Toulouse, France}
\author{P. Parneix}
\affiliation{Laboratoire de Photophysique Mol\'eculaire, B\^at. 210, Universit\'e
Paris-Sud, F91405 Orsay cedex, France.}
\begin{abstract}
Unimolecular evaporation in rotating atomic clusters is investigated using
phase space theory (PST) and molecular dynamics simulations. The
rotational densities of states are calculated in the sphere$+$atom
approximation, and analytical expressions are given for a radial interaction potential
with the form
$-C/r^p$. The vibrational densities of states are calculated using Monte Carlo
simulations, and the average radial potential at finite temperature is obtained
using a recent extension of the multiple range random-walk algorithm. These
ideas are tested on simple argon clusters modelled with the Lennard-Jones
interaction potential, at several total energies and angular momenta of the parent
cluster. Our results show that PST successfully reproduces the
simulation data, not only the average KER but its probability distribution, for
dissociations from LJ$_{14}$, for which the product cluster can effectively be
considered as spherical. Even for dissociations from the nonspherical
LJ$_8$, simulation results remain very close to the
predictions of the statistical theory.
\end{abstract}
\maketitle

\section{Introduction}

Fragmentation in finite systems offers a convenient way to investigate their
physical and chemical properties. In this respect atomic and molecular clusters
have received a great deal of attention, and experimental measurements of
structural\cite{ray89,brech89,hwang90,wei90,smith92,xu93,hild98} or
electronic\cite{deheer87,brech92} data have been reported using unimolecular
dissociation analyses. In particular, the relative stability of a cluster is
commonly characterized by its dissociation energies, giving rise to the well
known ``magic numbers'' in the mass spectra.\cite{tpmartin} Recently,
fragmentation has been employed to probe thermodynamical properties on a more
global scale, with a focus at phase transitions. Haberland and coworkers have
used photoabsorption induced fragmentation to extract caloric curves in charged
sodium clusters accross the solid-liquid phase change.\cite{schmidt97} They
have also extended their measurements to probe the liquid-vapor phase
change.\cite{schmidt01} At the same time, Gobet and coworkers used
event-by-event data analyses of multifragmentation in H$_3^+$(H$_2$)$_m$
clusters induced by collisions with a helium target,\cite{gobet02} showing a
small backbending in the caloric curve. Br\'echignac {\em et al.} also found
some evidences of the liquid-gas transition in small strontium clusters from
the shape of the kinetic energy release distribution subsequent to photoexcitation.\cite{brech02}

The possible correlations between statistical fragmentation and phase
transitions have found theoretical supports in cluster
physics,\cite{wa,parneix1,parneix2,calvo01} but also in nuclear
physics.\cite{gross90}
Decaying nuclei resulting from collisions\cite{pochodzalla95} typically show
features due to a very large energy deposit, where multiple fast fragments are
emitted on a short time scale. In this case, the main concern is to
characterize the distribution of the fragments, and the size of the remaining
droplet. In suitable situations, Fisher's formula\cite{fisher67} gives a
correct account of the mass distribution measured in experiments.

Atomic clusters are usually treated much more gently, by adding a small amount
of excitation energy. Only a very few atoms undergo dissociation, and the
evaporative process can take place over long time scales. For example, large
weakly bound rare-gas clusters can exhibit extremely small rate constants if
their excitation energy lies not far above the dissociation threshold, because the time
required for the excitation energy to be located on the few dissociative modes
rises sharply as the cluster size increases. In these systems, one is more
interested by a complete characterization of the evaporation event itself, with
a single ejected atom involved. Two observables carry most of the useful
information, namely the dissociation rate and the kinetic energy release (KER)
distribution. Weerasinghe and Amar (WA)\cite{wa} theoretically investigated
in great details the evaporation process in small argon clusters. Their results
show that the evaporation rate, and even more the average KER, can be used as a
probe of the solidlike-liquidlike phase change in the parent cluster.\cite{wa}
To achieve this result, they compared various statistical theories of
unimolecular dissociation to the outcome of molecular dynamics (MD)
simulations, below the energy range where MD becomes prohibitive. One of their
conclusions is that phase space theory (PST), in the sense of Chesnavitch and
Bowers,\cite{cb} is able to describe accurately the full evaporation statistics
in Ar$_n$ clusters, while simpler theories such as the Rice-Ramsperger-Kassel
(RRK) model\cite{rrk} or the Weisskopf-Engelking formula\cite{weissk,engelk}
only produce correct orders of magnitude.\cite{wa} Some important features,
including the nonlinear variation of average KER with increasing excitation
energy, are completely absent from the predictions of these approximate models.
Comparable methods have been applied by Peslherbe and Hase to the dissociation
in small aluminium clusters,\cite{hase} leading to similar conclusions.

Br\'echignac and coworkers recently reported time-of-flight mass spectrometry
measurements of evaporating Na$_n^+$ clusters.\cite{brech01} A careful
interpretation of these results necessitated the partitioning of the
translational and rotational kinetic energy released, because only the former
is actually measured. The possible angular momentum of the parent cluster is a
problem, since rotation can strongly alter the evaporation dynamics, hence the
statistical observables. Up to now, only few theoretical works have been
devoted to the dynamics of rotating clusters. Structural properties and angular
momentum driven isomerizations were first investigated by Jellinek and
Li.\cite{jelli89a,jelli89b,jelli90} Using simple statistical theories, Miller
and Wales further investigated static and evaporation properties on the
effective rovibrational potential energy surface (PES).\cite{wales1} The
influence of angular momentum on cluster thermodynamics\cite{mcrot} and
chaotic dynamics\cite{calvo00y} also received some attention. Evaporation in
rotating clusters had been previously investigated by Stace using simple
models in the framework of phase space theory.\cite{stace91} The
calculations made by this author showed that angular momentum tends to increase
in small clusters after evaporation (rotational heating), while it tends to
decrease in large clusters (rotational cooling). These effects have been partly
observed in the MD simulations performed by Weerasinghe and Amar.\cite{wa}

As seen from the success of phase space theory to describe evaporation in
nonrotating clusters,\cite{wa} it is highly desirable to extend this work to
the case of finite angular momenta. This is the goal of the present paper. In
the next section, we give the basic PST formalism needed to calculate the rotational
density of states, in the sphere$+$atom approximation. Exact results are obtained
for a radial interaction potential having the form $-C/r^p$, and we provide further
details about the numerical implementation of the method in the more general
case of a $-C/(r-r_0)^p$ interaction. The other
computational ingredients include the estimation of the vibrational density of
states as well as the radial interaction potential. We also carry out some MD simulations
to be used as a benchmark for testing the predictions of PST. Application is
made in Sec.~\ref{sec:res} to the evaporation in Ar$_{14}$ and Ar$_8$,
modeled using the common Lennard-Jones potential. We finally summarize and
conclude in Sec.~\ref{sec:ccl}.

\section{Phase space theory}
\label{sec:pst}

In this Section, we work out the main expressions for the distribution of
kinetic energy released during evaporation of rotating polyatomic molecules.
Conservation of angular momentum $J$ is rigorously included in the phase
space theory.\cite{cb,pechukas65,klots71} This is particularly important when
treating rotating systems with prescribed values of $J$. Additionally, PST is
built upon the hypothesis of a loose transition state, i.e. the products are
the transition state. In rotating clusters, the centrifugal barrier
previously explicitely considered by Miller and Wales,\cite{wales1} is
naturally accounted for in PST.

Here we consider a parent cluster characterized by a rotational angular
momentum $J$ and a total rovibrational energy $E$. We denote by $J_r$ the
rotational angular momentum of the subcluster (product) after dissociation.
Following WA and Jarrold,\cite{jarrold} the probability of finding a
dissociation event with $\etr$ kinetic energy released is given within $d\etr$
by:
\beq
P(\etr,E,J) = R(\etr,E,J)\left/ \int_{\etr^{\rm min}}^{E-E_0^{(J)}} R(\etr,E,J)
d\etr\right.,
\label{eq:peej}
\eeq
with the differential rate $R(\etr,E,J)$
\beq
R(\etr,E,J)=R_0 \frac{\Omega_n^{(J)}(E-E_0^{(J)}-\etr)\Gamma(\etr,J)}
{\Omega_{n+1}^{(J)}(E-E_r)}.
\label{eq:reej}
\eeq
In the latter equation, $R_0$ is a constant factor that accounts for channel
and rotational degeneracies.\cite{wa} $E_r$ is the rotational energy of the parent 
cluster. $\Omega_{n+1}^{(J)}$ and $\Omega_n^{(J)}$
are the vibrational densities of states (VDOS) at angular momentum $J$
of the parent and product clusters, respectively. $\Gamma$
is the rotational density of states (RDOS) of the fragments.
In these notations, we have implicitely assumed that
both densities of states of the product cluster depend on $J$, but depend only weakly
on $J_r$. In the same line of ideas, we consider that the clusters are large
enough so that the energy difference $E_0^{(J)}$ between the potential energy
minima of the parent and product clusters can be taken at the same value of
$J$ for both clusters.
The knowledge of the differential rate function $R$ readily leads to the
average kinetic energy released:
\beq
\langle \etr \rangle = \int_{\etr^{\rm min}}^{E-E_0^{(J)}} \etr P(\etr,E,J)
d\etr.
\label{eq:avetr}
\eeq
The calculation of $P(\etr,E,J)$ and $\langle\etr\rangle$ requires one to
compute both $\Omega$ and $\Gamma$ for the {\em product}\/ cluster, but neither
the constant $R_0$ nor the VDOS for the parent cluster. The absolute rate
constant obtained from integrating $R(\etr,E,J)$ with respect to $\epsilon_{tr}$ thus requires a more
substantial effort than any data related to the KER. In the remainder of this
section, we focus on the rotational density of states. The vibrational
quantities are relatively easy to compute, and they will be dealt with later.

The calculation of the $\Gamma(\etr,J)$ function is directly linked to the
energetics and angular momentum constraints during dissociation. Our main
assumption will be to treat the evaporative system LJ$_{n+1}\to$LJ$_n+$LJ as
well represented by a sphere$+$atom model. Then the product cluster LJ$_n$ has
a unique rotational constant $B$. In this case, Chesnavitch and Bowers have
shown that the rotational density of states can be calculated as\cite{cb}
\beq
\Gamma(\etr,J)=\int\!\!\!\!\int_{\cal S} \Gamma(\varepsilon_r^*,J_r)dJ_rdL,
\label{eq:gamma4}
\eeq
where $\varepsilon_r^*$ stands for the upper limit of the rotational energy,
$L$ for the orbital angular momentum of the products. The integration is
carried out in the ($J_r,L$) plane with the boundaries ${\cal S}$ discussed
below. In the sphere$+$atom case, $\Gamma(\varepsilon_r^*,J_r)$ simply equals
$2J_r$,\cite{jarrold} hence the problem is reduced to finding the expressions
for the boundaries ${\cal S}$. We first consider the case of a radial
dissociation potential $V(r)$ given by $V(r)=-C/r^p$, with $p$ greater than 2.

\begin{figure*}[htb]
\setlength{\epsfxsize}{17cm}
\leavevmode\epsffile{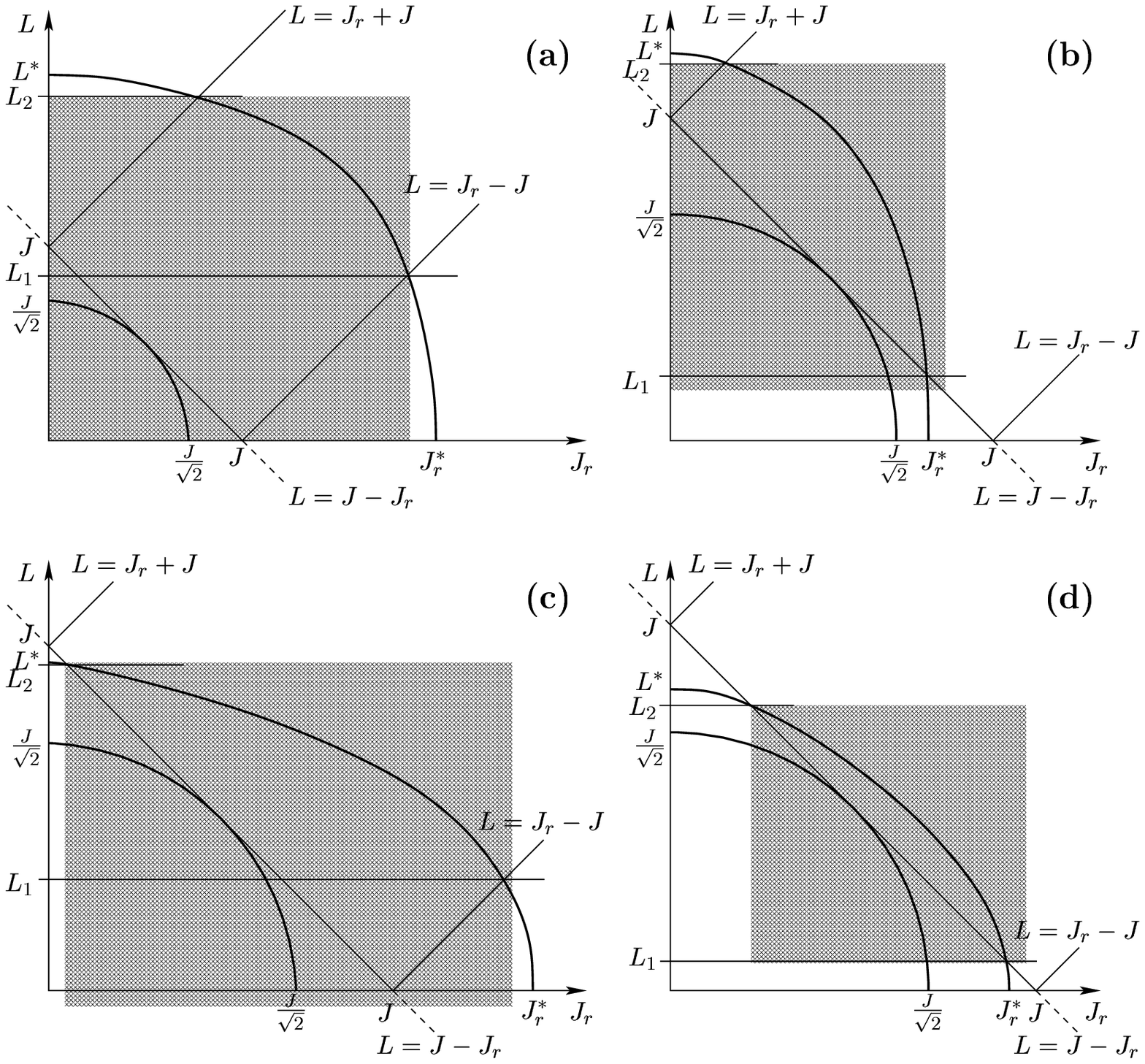}
\caption{Schematic representation of the $(L,J_r)$ integration plane, denoted
by ${\cal S}$ in the text.
(a) $J<L^*$ and $J<J_r^*$; (b) $J_r^*\leq J<L^*$; (c) $L^*\leq J < J_r^*$;
and (d) $J>L^*$ and $J>J_r^*$. The outer boundary is defined by
the curve ${\cal C}$ (see text), the inner boundary is the circle with radius
$J$ centered at (0,0).}
\label{fig:schema}
\end{figure*}

The first boundary on ${\cal S}$ is given by the constraint on the kinetic
energy of the dissociating atom. The centrifugal barrier $\varepsilon^\dagger$
that must be overcome is located at $r=r^\dagger$ such that $V_L(r)=V(r)+L^2/
2\mu r^2$ is maximum. This yields
\beq
\varepsilon^\dagger = L^{2p/(p-2)}/\Lambda_p,
\label{eq:epsd}
\eeq
with the notation
\beq
\Lambda_p= \frac{2}{p-2}C^{2/(p-2)}\left( \frac{\mu p}{\hbar^2}\right)^{p/(p-2)}.
\label{eq:lambda}
\eeq
For the atom to actually dissociate, its kinetic energy must be positive at
the barrier, which is expressed as
\beq
BJ_r^2 + L^{2p/(p-2)}/\Lambda_p \leq \etr.
\label{eq:cond7}
\eeq
The second boundary on ${\cal S}$ comes from the conservation of angular
momentum, ${\vec J} = {\vec J}_r + {\vec L}$, or
\beq
|J_r - L| \leq J \leq J_r+L.
\label{eq:cond8}
\eeq
The conditions (\ref{eq:cond7}) and (\ref{eq:cond8}) define lower and upper
bounds for $J_r$ at each value of $L$, denoted as $J_r^{\rm min}(L)$ and
$J_r^{\rm max}(L)$, respectively. Integration over the contour ${\cal S}$ can
be formally carried out:
\beq
\Gamma(\etr,J) = \int_{L_{\rm min}}^{L_{\rm max}} \left[ (J_r^{\rm max}(L))^2
-(J_r^{\rm min}(L))^2\right] dL,
\label{eq:gamma9}
\eeq
where we have introduced the lower and upper bounds for the integration on $L$,
namely $L_{\rm min}$ and $L_{\rm max}$. Let now ${\cal C}$ be the set of
$(J_r,L)$ points that fulfill the equation $\etr=L^{2p/(p-2)}/\Lambda_p
+BJ_r^2$. Let also $J_r^*$ and $L^*$ be the intersection points of ${\cal C}$
with the abscissa and ordinates axes, respectively. We find that
\beq
J_r^* = (\etr/B)^{1/2}  \mbox{~~~and~~~} L^* = (\Lambda_p\etr)^{(p-2)/2p}.
\label{eq:jrl}
\eeq
Four different cases must be treated separately, depending on whether the values of
$J_r^*$ and $L^*$ are smaller or larger than the initial angular momentum $J$.
These four cases are depicted in Fig.~\ref{fig:schema}. They correspond to the
following conditions:
\begin{eqnarray*}
{\rm (a)} && J<J_r^*;~ J<L^*, \\
{\rm (b)} && J_r^*\leq J<L^*, \\
{\rm (c)} && L^*\leq J<J_r^*, \\
{\rm (d)} && J_r^*\leq J;~ L^*\leq J.
\end{eqnarray*}
In cases (a) and (c), integration starts at $L_{\rm min}=0$. In cases (b) and
(d), $L_{\rm min}$ is determined by the intersection of ${\cal C}$ with
$L=J-J_r$. We denote this point by $L_1$:
\beq
\left\{ \begin{array}{l}
L=J-J_r,\\
L^{2p/(p-2)}/\Lambda_p + BJ_r^2=\etr. \end{array}\right.
\label{eq:l1}
\eeq
In (a) and (b), $L_{\rm max}=L_2$ is obtained at the unique intersection of
${\cal C}$ with $L=J+J_r$:
\beq
\left\{ \begin{array}{l}
L=J+J_r,\\
L^{2p/(p-2)}/\Lambda_p + BJ_r^2=\etr. \end{array}\right.
\label{eq:l2}
\eeq
Finally, the upper bound $L_{\rm max}=L_2$ is given in cases (c) and (d) by the
intersection of ${\cal C}$ with $L=J-J_r$, equation (\ref{eq:l1}) above. Thus
in (d) the two extremal values are solutions of the same equation.

At any value of $L$ in the range $L_{\rm min}\leq L\leq L_{\rm max}$, the
lower and upper values of $J_r(L)$ also depend on the conditions (a--d). For
example, in case (a) one must distinguish between three subcases, namely
$0\leq L\leq L_1$, $L_1\leq L\leq J$, and $J\leq J\leq L_2$, where $L_1$
denotes the intersection of ${\cal C}$ with $L=J_r-J$. In the range $0\leq L
\leq J$, $J_r^{\rm min}$ is equal to $J-L$. In the range $J\leq L\leq L_2$,
$J_r^{\rm min}(L)=L-J$. The upper bound $J_r^{\rm max}$ is given by $J_r^{\rm
max}=J+L$ for $0\leq L\leq L_1$, and by $J_r^{\rm max}(L)=[\etr - L^{2p/(p-2)}
/\Lambda_p]/B$ for $L_1\leq L\leq L_2$.

After some algebra, integration over the
boundary ${\cal S}$ leads to the total rotational density of states:
\begin{eqnarray}
\Gamma(\etr,J) &=& (L_2-L_1)(\etr/B - J^2) \nonumber \\
&& - \frac{1}{\Lambda_p B}
\frac{p-2}{3p-2}\left( L_2^\frac{3p-2}{p-2}
-L_1^\frac{3p-2}{p-2}\right) \nonumber \\
&&+ J(L_2^2+L_1^2) - (L_2^3-L_1^3)/3.
\label{eq:gam13}
\end{eqnarray}
We will not discuss the three other cases (b--d) in details, and we only
provide below the final results. In cases (a) and (c), the RDOS is given by
Eq.~(\ref{eq:gam13}) above. In cases (b) and (d), it is expressed by
\begin{eqnarray}
\Gamma(\etr,J) &=& (L_2-L_1)(\etr/B - J^2) \nonumber \\
&&- \frac{1}{\Lambda_p B} \frac{p-2}{3p-2}\left(
L_2^\frac{3p-2}{p-2} - L_1^\frac{3p-2}{p-2}\right)
\nonumber \\
&& + J(L_2^2-L_1^2) - (L_2^3-L_1^3)/3,
\label{eq:gam14}
\end{eqnarray}
which only differs from Eq.~(\ref{eq:gam13}) by the quantity $2JL_1^2$. 

We have not yet discussed the lower bound of integration on $\etr$ in
Eqn.~(\ref{eq:peej}) and (\ref{eq:avetr}), denoted as $\etr^{\rm min}$. This
limit occurs when the curve ${\cal C}$ is tangent to the line $L=J-J_r$. This
condition can be cast into an equation in $J_r$ only:
\beq
\frac{p-2}{p}B\Lambda_p J_r(J-J_r)^\frac{2+p}{2-p} = 1,
\label{eq:tangent}
\eeq
which can be easily shown to have a unique solution in the range $0\leq J_r
\leq J$. The value of $\etr^{\rm min}$ follows from substitution in $({\cal
C})$ and $L=J-J_r$. In the case of a van der Waals dispersion interaction,
$V(r)=-C/r^6$, the appropriate values for the rotational density of states
are given exactly for $p=6$ by
\begin{eqnarray}
\Gamma(\etr,J)&=&(L_2-L_1)(\etr/B-J^2) -\frac{L_2^4-L_1^4}{4\Lambda_6 B}\nonumber \\
&&+J(L_2^2 \pm L_1^2) -(L_2^3-L_1^3)/3
\label{eq:gam13p}
\end{eqnarray}
where the plus sign stands in cases (a) and (c), and the minus sign stands in (b) and (d).
In the case $p=6$, an anaytical expression for $\etr^{\rm min}$ can also be found
\begin{eqnarray}
\etr^{\rm min} &=& BJ^2+\frac{2}{3}B^2\Lambda_6 J + \frac{2}{27}B^3\Lambda_6^2
\nonumber\\
&&- \left(\frac{2}{27}B^3\Lambda_6^2 + \frac{4}{9}B^2\Lambda_6 J\right)\left( 1 +
\frac{6J}{B\Lambda_6}\right)^{1/2},
\label{eq:etrmin}
\end{eqnarray}
which is given in the low $J$ regime by $\etr^{\rm min} =J^3/\Lambda_6$,
up to the fourth order in $J$.
The numerical implementation of the above formulas is straightforward. At fixed
total energy $E$ and angular momentum $J$ of the parent cluster, one must first
calculate the lower and upper limits $\etr^{\rm min}$ and $\etr^{\rm max}=E
-E_0^{(J)}$. For each value of $\etr$ in this range, equations (\ref{eq:l1})
and (\ref{eq:l2}) must be solved numerically to give $L_1$ and $L_2$, and the
rotational density of states is calculated using Eqn.~(\ref{eq:gam13}) and
(\ref{eq:gam14}) above.

All the previous formalism has been derived by assuming an interaction
potential with the form $V(r)=-C/r^p$. It turns out that this expression does
not give a very good account of the finite extension of the cluster, and that
a better representation of the atom-cluster interaction is provided by $V(r)=
-C/(r-r_0)^p$, with $r_0>0$. In this case, and more generally for an arbitrary
form of $V(r)$, the computation of the rotational density of states must be
carried out numerically. Firstly, for a series of $L$, the location $r^*(L)$ 
of the centrifugal barrier is obtained by solving
\beq
\left. \frac{\partial V}{\partial r}\right|_{r=r^*} = \frac{L^2}{\mu (r^*)^3}.
\label{eq:rstar}
\eeq
One then deduces the height $\varepsilon^\dagger(L)$ of the barrier:
\beq
\varepsilon^\dagger(L)=\frac{L^2}{2\mu (r^*)^2} - V (r^*).
\label{eq:epsd2}
\eeq
At a given $\etr$, the integration boundaries ${\cal S}$ become
\beq
\left\{ \begin{array}{l}
\varepsilon^\dagger(L)+BJ_r^2 \leq \etr, \\
|J_r-L|\leq J\leq J_r+L \end{array} \right.
\label{eq:snum}
\eeq
and the limits $L_1$ and $L_2$ are still given by equations (\ref{eq:reej})
and (\ref{eq:avetr}) after replacing $L^{2p/(p-2)}/\Lambda$ by
$\varepsilon^\dagger(L)$. While $(J_r^{\rm min})^2$ is still equal to $(J-L)^2$,
$(J_r^{\rm max})^2$ is now obtained from $[\etr - \varepsilon^\dagger]/B$ in
the range $L_1\leq L\leq L_2$. The integration of $\Gamma(\etr,J)$ must be
done numerically, after estimating $\etr^{\rm min}$ from the tangency condition
of $\etr=\varepsilon^\dagger+BJ_r^2$ with $L=J_r-J$.

\section{Computational procedure}
\label{sec:ing}

\subsection{Vibrational density of states}

The vibrational densities of states $\Omega^{(J)}(E)$ depend implicitely on the
total angular momentum $\vec J$\/ of the cluster. Actually this dependence acts
by two ways.\cite{mcrot} Firstly, the centrifugal effects perturb the potential
energy surface into an effective, rovibrational surface.\cite{jelli89a,wales1}
Secondlya, the conservation of the vector $\vec J$ adds an extra geometrical
weight in the configurational density of states or in the partition function.
This weight is given explicitely by $1/\sqrt{{\rm det}\, {\bf I}}$, where
${\bf I}$ is the inertia tensor at the current configuration.\cite{mcrot,deti}
The angular momentum is naturally conserved in constant energy
molecular dynamics simulations, as well as in constant temperature
Nos\'e-Hoover schemes at $\vec J=\vec 0$, but not in conventional Monte Carlo
simulations. Therefore, some differences between the MD and MC procedures may
arise when this weight may span several orders of magnitude, as in the case of
Ar$_3^+$, which is linear in its ground state geometry.\cite{galindez}

To calculate $\Omega^{(J)}(E)$ for several values of $J$, we have used the
Monte Carlo method proposed in Ref.~\onlinecite{mcrot}, further improved with the
parallel tempering accelerating scheme.\cite{ptmc} The multiple histogram
method\cite{histo} was then used to estimate the configurational densities of
states. In turn, the total vibrational densities of states were obtained from
the configurational densities by a simple convolution product. 

The present results were checked by performing additional molecular
dynamics simulations. The histogram analysis\cite{calvo1} showed a good
agreement between the MC and the MD calculation.
We have also checked the physical relevance of the calculation by computing other
related thermodynamical observables, such as the canonical heat capacity.
At low values of $\vec J$, the melting temperature was seen to roughly decrease
with $\vec J$ as $J^2$, in agreement with previous works.\cite{mcrot}

\subsection{Radial potential}

In a first approach, the dissociation potential $V(r)$ felt by an atom
leaving the $n$-atom Lennard-Jones cluster can be approximated by its
asymptotic form $-C_6^{(n)}/r^6$. At very large distances $r$, the $C_6^{(n)}$
parameter is given by 4$n\varepsilon\sigma^6$ LJ units.
However, at intermediate distances,
where the centrifugal barrier is likely to be located, the finite extent of
the cluster induces significant deviations of the average potential. A simple
approach to this problem is to consider a continuous homogeneous distribution
of Lennard-Jones centers inside a sphere of radius $R\propto n^{1/3}$. For
large sizes, this leads to the Gspann-Vollmar potential $V_n$:\cite{gspann}
\begin{eqnarray}
V_n(r)&=&C_{12}\frac{r^6+21r^4r_0^2/5+3r^2r_0^4+r_0^6/3}{(r^2-r_0^2)^9}
\nonumber \\
&&-\frac{C_6}{(r^2-r_0^2)^3},
\label{eq:gspann}
\end{eqnarray}
where $C_{12}$, $C_6$ and $r_0$ are size-dependent and given by
\begin{eqnarray}
C_{12}&=&4n\varepsilon \sigma^{12}; \nonumber \\
C_6&=&4n\varepsilon \sigma^6 ; \label{eq:paragspann} \\
r_0&=&(3/p\pi\rho)^{1/3}[n^{1/3}-1].
\end{eqnarray}
In the above equation, $\rho$ is the atomic density in the solid state.
The Gspann-Vollmar potential was built in a similar way as the Girifalco
potential describing the interaction between C$_{60}$
molecules.\cite{girifalco} It is not appropriate for medium-size, nonspherical
clusters, or for intermediate distances $r$, where the continuous approximation
would break down. In addition, because the cluster is thermalized at a finite
temperature,
the atomic fluctuations may induce some changes in the average potential felt
by the tagged distant atom. Using constraint dynamics, Weerasinghe and
Amar\cite{wa} showed that the simple $-C_6/r^6$ form was not fully
appropriate to describe the atom-cluster interaction, and that a much better
fit was obtained using the $-C_6/(r-r_0)^6$ form. The Gspann and Vollmar
results also suggest that a $-C_6/(r^2-r_0^2)^3$ form could also be used.
We have carried out some constrained Monte Carlo simulations at finite
temperature, by keeping the external atom at a fixed distance $r$ from the
cluster center of mass. The temperature effects were not investigated in the
paper by Weerasinghe and Amar, and we have chosen to perform the calculation
at low (0.01 LJ units) and high $T$, namely $T=0.2$ for LJ$_8$ and $T=0.3$
for LJ$_{14}$. These values are close to the melting points of the two
clusters, above which evaporation takes place in a sub-nanosecond time scale.

\begin{figure}[htb]
\setlength{\epsfxsize}{9cm}
\centerline{\epsffile{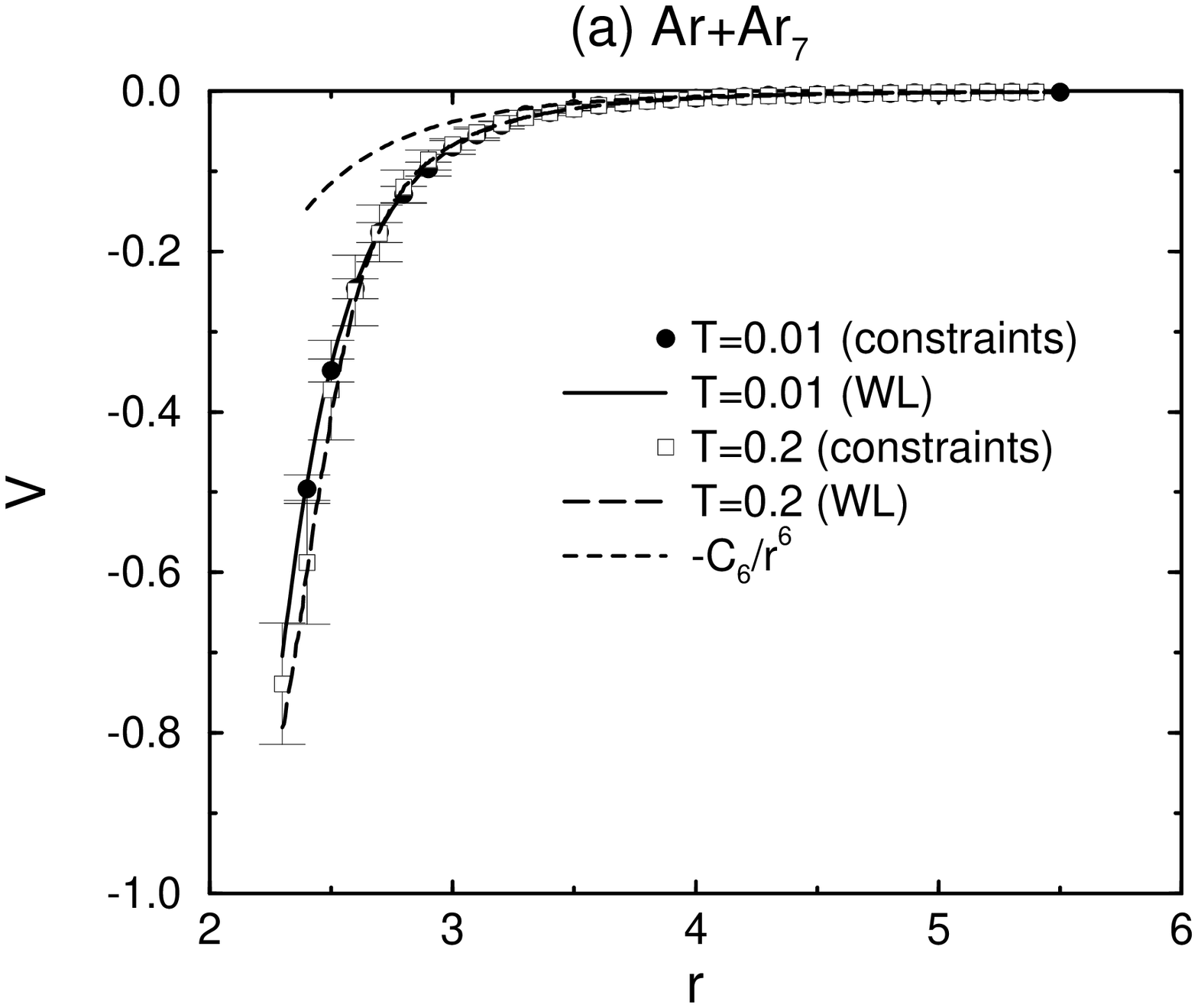}}
\setlength{\epsfxsize}{9cm}
\centerline{\epsffile{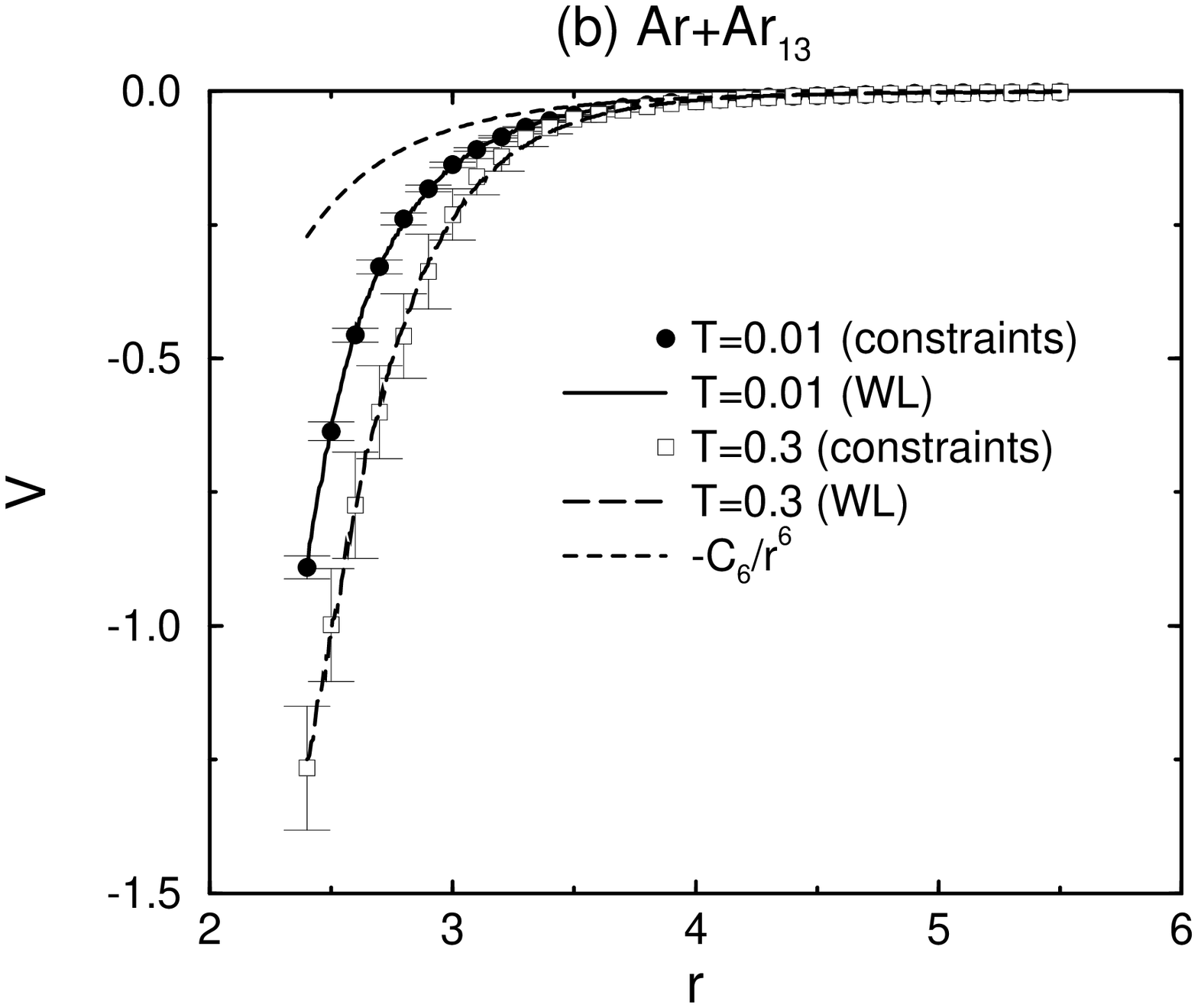}}
\caption{Atom-cluster radial potential in the reaction LJ$_{n+1}\to$LJ$_n
+$LJ. The symbols are from constrained MC simulations, the solid and long
dashed lines are the result of the Wang-Landau (WL) multicanonical
reweighting scheme. The asymptotic law $-C_6/r^6$, with $C_6=4n$ LJ units,
is also drawn as a dashed line.
The data are plotted for two temperatures in each case. (a) $n=7$; (b) $n=13$.}
\label{fig:vradial}
\end{figure}

As an alternative to the MC simulations with constraints, we have calculated
the finite temperature dissociation potential using the recently proposed
multiple range random walk algorithm by Wang and Landau.\cite{wl} This method
has been straightforwardly extended to the computation of effective potentials
and potentials of mean forces,\cite{fcwl} and consists of performing a
Metropolis Monte Carlo simulation using the following acceptance rule:
\begin{eqnarray}
&{\rm acc}({\bf R}_{\rm old}\to {\bf R}_{\rm new})= \nonumber \\
&\min\left[ 1,
\frac{g(r_{\rm old})}{g(r_{\rm new})}\exp\{-\beta[V({\bf R}_{\rm new})-
V({\bf R}_{\rm old})]\}\right],
\label{eq:wl1}
\end{eqnarray}
where $\beta=1/k_BT$,
${\bf R}_{\rm old}$ and ${\bf R}_{\rm new}$ are two successive points in
the configuration space, $r_{\rm old}$ and $r_{\rm new}$ the corresponding
atom-cluster distances, respectively. $g(r)$ is a weight function, initially
set to 1 in the entire range of $r$, which evolves dynamically along the MC
simulation by the operation $g(r)\to f\times g(r)$ after the distance $r$ has
been visited. The constant factor $f$ is initially set to 2, and gradually
decreases to 1 after a given number of Monte Carlo steps. After several
iterations of this process, the function $\Gamma(r)=-\beta^{-1}\ln g(r)$
converges to the potential of mean force (PMF) $W(r)$:\cite{fcwl}
\begin{equation}
\Gamma(r)\to W(r)=-\beta^{-1}\ln p(r),
\label{eq:gammar}
\end{equation}
where $p(r)$ is the probability distribution of finding the atom at the
distance $r$ from the cluster, given by the canonical average $p(r_0)=
\langle \delta[r_0 - r({\bf R})]\rangle$.

Once  the PMF  is  known, it  can  be subsequently  used  in a  biased
multicanonical simulation  to sample  the  entire  range of  distances  in a
uniform way. For this we replace the potential $V$ by $V+W$. The  average
potential $V(r)$  felt by the  atom at distance
$r$ from the cluster center of  mass is given by the usual reweighting
formulas.\cite{allen} The Wang-Landau scheme allows one to compute the
potential $V$ over a continuous range of $r$, instead of only a
small set when using constraint dynamics.

We have represented in Fig.~\ref{fig:vradial} the effective potential
calculated from constrained MC simulations and from multicanonical
simulations for the two sizes $n+1=8$ and $n+1=14$, at low and high
temperatures. In each case, we observe a very good agreement between
the two methods, suggesting that the Wang-Landau/multicanonical scheme
can yield accurate average potentials over continuous range of distances.
For the two clusters, the interaction between the external atom and the
cluster is stronger at higher temperature. This can be explained by the
larger ``volume'' of the cluster in its liquidlike state with respect to
its solidlike low temperature value. Hence the apparent extent of the
cluster is larger, and the potential is larger in modulus.
As can be seen in the two pannels of Fig.~\ref{fig:vradial}, the asymptotic
$-C_6/r^6$ form is not appropriate at intermediate distances when taking
$C_6=4n\varepsilon\sigma^6$. As noted by Weerasinghe and Amar,\cite{wa}
setting this constraint on $C_6$ free does not improve the
behavior of $V(r)$ much. A better fit is obtained with the expression
$V(r)=-C_6/(r-r_0)^6$. The values of $C_6$ and $r_0$ as a function of
cluster size and temperature are given in Table~\ref{tab:c6r0}.

\begin{table}[htb] 
\caption{Fitting parameters $C_6$ and $r_0$ of the average atom-cluster LJ
potential $V(r)=-C_6/(r-r_0)^6$, and average rotational constant $B$, for LJ$_n$ clusters
at different temperatures.}
\label{tab:c6r0}
\begin{tabular}{|c|r@{.}l|r@{.}l|r@{.}l|r@{.}l|}
\colrule
Cluster & \multicolumn{2}{c|}{Temperature}& \multicolumn{2}{c|}{$C_6$} & \multicolumn{2}{c|}{$r_0$}& \multicolumn{2}{c|}{$B$}\\
size $n$& \multicolumn{2}{c|}{($\varepsilon/k_B)$} & \multicolumn{2}{c|}{($\varepsilon\sigma^6$)}& \multicolumn{2}{c|}{($\sigma$)}& \multicolumn{2}{c|}{($m\sigma^2$)} \\
\colrule
7 & ~~~~~~0&01 & 8&070 & 0&809& 0&150 \\
7 & 0&20 & 8&728 & 0&802& 0&135 \\
\colrule
13 & ~~~~~~0&01 & 19&149 & 0&733 & 0&0534 \\
13 & 0&30 & 62&353 & 0&499 & 0&0444 \\
\colrule
\end{tabular}
\end{table}

Temperature effects are weak on the smaller cluster, as both $C_6$ and $r_0$
remains nearly constant. This is probably due to the fact that, upon melting,
the non spherical LJ$_7$ cluster does not really enlarge, but instead visits
other (still non spherical) isomers. On the other hand, a significant change
is seen on the parameters for the much more spherical $n=13$. At large
temperature, the values we
get are found to yield a very similar average potential than the one found
by Weerasinghe and Amar.\cite{wa} However, because these authors employed
constant energy MD simulations and because they did not provide the total
energy used for this cluster, we cannot reliably compare our results with
theirs.

We now have all the ingredients required for the PST calculations. In order
to assess or question the quality of the statistical theory, we need
to carry out simulations of the actual evaporation process at finite
angular momentum.

\subsection{Molecular dynamics simulation of the evaporation dynamics}

For each size and for each value of angular momentum, we have considered
a set of 5000 molecular dynamics trajectories. The average kinetic energy
release has been analysed after each trajectory ending into an evaporation
event. The two contributions of the KER were evaluated, namely the
rotational part of the  product LJ$_n$ subcluster, and the translational
contribution of the external atom undergoing evaporation.
The instant of evaporation was considered as the last time at which the
radial velocity of the atom was negative.\cite{wa}
By varying the total angular momentum between $J=0$ and $J \approx 5$ LJ
units,\cite{conversion} we have performed a
systematic study of the effects of rotation on the evaporation process.

\section{Results and discussion}
\label{sec:res}

In this paper, we will focus
on the energetics of the unimolecular process, rather than on the absolute
evaporation constant.
This choice is mainly guided from
previous studies where the relationship between phase transition in the
product clusters and the evaporation statistics was most clearly evidenced
on the kinetic energy released.\cite{wa,parneix1,parneix2}
Two different unimolecular reactions
have been considered, involving the nearly spherical LJ$_{13}$ product
and the nonspherical LJ$_7$ system. This latter cluster has an ellipsoidal
symmetry with inertia momenta in the ratio (0.64,0.64,1) at $T=0$. The two reactions
studied here are thus LJ$_{14}\to$LJ$_{13}+$LJ\ and LJ$_{8}\to$LJ$_{7}+$LJ.

\subsection{\boldmath LJ$_{14}\to$LJ$_{13}+$LJ\unboldmath}

LJ$_{13}$ is a magic cluster and shows enhanced melting point and latent
heat of melting with respect to its immediate neighbors LJ$_{12}$ and
LJ$_{14}$. In addition to its spherical shape, this cluster provides a good
candidate for investigating the melting transition as seen from its evaporation
observables.

\begin{figure}[htb]
\setlength{\epsfxsize}{9cm}
\epsffile{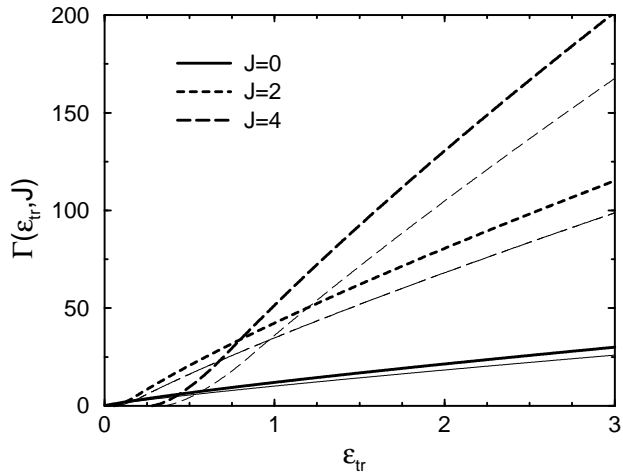}
\caption{Rotational density of states $\Gamma (\epsilon_{tr},J)$
as a function of $\etr$ for 3 values of $J$ in the unimolecular dissociation
of LJ$_{14}$. The curves plotted are the predictions of PST using
the simulated radial potential at $T=0.3$ (thick lines) or the simpler $-C_6/r^6$
potential with $C_6=52$ (thin lines).}
\label{fig:gamma13}
\end{figure}

To apply the PST formalism, we first have to calculate
the rotational density of states $\Gamma (\etr,J)$ for different
values of the angular momentum $J$. As explained in Section~\ref{sec:pst},
$\Gamma (\etr,J)$\ is sensitive
to the radial potential $V(r)$ felt by the dissociating atom. It is also
sensitive to the rotational constant $B$ of the product LJ$_{13}$
cluster. In Fig.~\ref{fig:gamma13}, $\Gamma (\etr,J)$ is plotted for $J$=0,
2, and 4 using the radial potential $-C_6/(r-r_0)^6$ and the rotational
constant calculated at moderate temperature, $T=0.3$, close to the melting
point. The parameters are given in Table~\ref{tab:c6r0}. For comparison, we
have also plotted the RDOS calculated using the simpler radial potential
$-C_6/r^6$, with $C_6=4n$ and the rotational constant at $T=0$. The latter
potential does not fully reproduce the simulated potential, but the
asymptotic limit (large $r$)
is known to be exact. Because the discrepancy between the
two radial potentials is quite large, the significant differences between the
predictions of PST in the rotational densities show that the centrifugal
barriers are indeed located within the range of distances plotted in
Fig.~\ref{fig:vradial}. The PST calculation using the analytical results at
$r_0=0$ notably underestimates the RDOS, by about 15\% for any $J$ and $\etr$.
The effects of a finite angular momentum in the parent cluster are strong.
The value of $\etr$ at which $\Gamma$ sharply increases, previously denoted as
$\etr^{\rm min}$, clearly changes with $J$. Moreover, the slope of the RDOS
also increases with $J$. As a consequence,
there is an order of magnitude increase at $\etr=3$ between $J=0$ and $J=4$.
This latter value can be considered as still moderate for LJ$_{14}$, as there is
only a small variation in the potential energy surface, and in the related
properties such as the heat capacity.\cite{wales1,mcrot}
In particular, the cluster
structure is only slightly perturbed, and spontaneous isomerization is
not expected to take place below about 15 LJ units.\cite{jelli89b}

\begin{figure}[htb]
\setlength{\epsfxsize}{9cm}
\epsffile{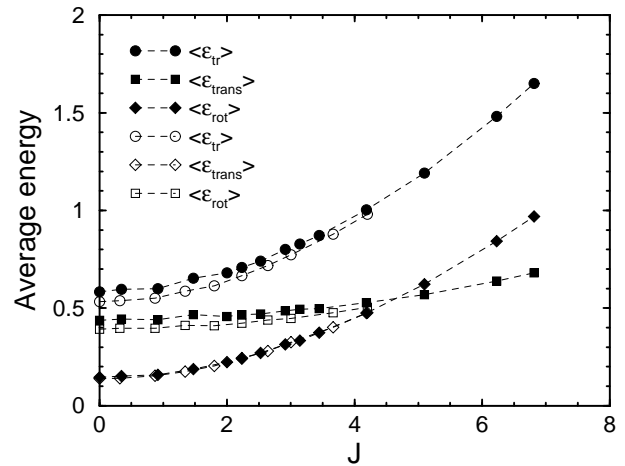}
\caption{Average translational ($\langle \varepsilon_{\rm trans}\rangle$, squares),
rotational ($\langle\varepsilon_{\rm rot}\rangle$, diamonds) and total ($\langle\etr\rangle$,
circles) kinetic energies released as a function of $J$ in the dissociation of LJ$_{14}$ at $E$=-26.
All results are from MD simulations. Full symbols correspond to $E=-26$; empty symbols are for $E=-29$.}
\label{fig:eav13}
\end{figure}

In Fig.~\ref{fig:eav13} we have represented the results of the molecular
dynamics simulations to be used as benchmarks for the present theoretical
analyses. The rotational, translational and total kinetic energy released
are plotted as a function of $J$ for two different total energies of the parent
cluster. The effect of $J$ on the energetics of
the dissociation reaction is mainly governed by the evolution of the rotational
contribution of the KER. Around $J=4$, this contribution becomes larger
than the translational energy. This latter contribution appears
almost constant up to $J\sim 4$ and slighty increases at higher $J$.  
In the range of energies considered here, the effect of internal energy is
weak. We observe a steady increase in the energies released during evaporation
as both angular momentum and internal energy increase, as intuition would
suggest.

\begin{figure}[htb]
\setlength{\epsfxsize}{9cm}
\epsffile{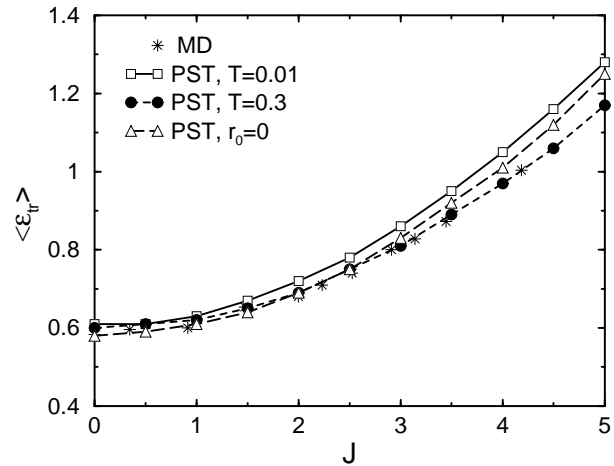}
\caption{Average total kinetic energy release in the dissociation of LJ$_{14}$
as a function of angular momentum $J$ at $E$=-26.
Comparison between MD simulation and phase space theory
using different radial potentials.}
\label{fig:etrav13_J}
\end{figure}

In the form detailed above, phase space theory only gives us access to the
total (translational$+$rotational) kinetic energy released. We have plotted in
Fig.~\ref{fig:etrav13_J} the variations of the average KER $\langle
\etr\rangle$ calculated from MD simulations as a function of $J$, and compared
them to the predictions of PST in the following three
approximations. The radial potential was either taken with $r_0=0$ or with
finite $r_0$, and in the latter case, two temperatures were taken,
corresponding to either the rigid case ($T=0.01$) or to the liquid case
($T=0.30$). The values of the parameters $r_0$, $C_6$ and $B$ implicitely
depend on these approximations. As can be seen from Fig.~\ref{fig:etrav13_J},
all three approximations perform quite well. Looking more closely, we notice
that the PST calculation at low temperature overestimates $\langle\etr\rangle$,
and that the approximation $r_0=0$ leads to a slightly diverging KER above $J\sim
3$. Actually, only the calculation at finite $r_0$ and temperature close to the
melting point remains quantitatively close to the simulation data in the entire
range of $E$. This is not
so surprising, because the simulation takes place at rather large internal
energies, where the cluster is in a liquidlike state. At low temperature, the average
rotational constant is larger (see Table~\ref{tab:c6r0}). Therefore the rotational
contribution to $\etr$ is overestimated, which explains the relatively high values of
$\langle\etr\rangle$ in Fig.~\ref{fig:etrav13_J}. Another possible cause could be the
less attractive interaction potential at this low temperature (see the lower panel of
Fig.~\ref{fig:vradial}), resulting in higher centrifugal barriers, hence a further shift
of $\etr$ to a larger value.

\begin{figure}[htb]
\setlength{\epsfxsize}{9cm}
\epsffile{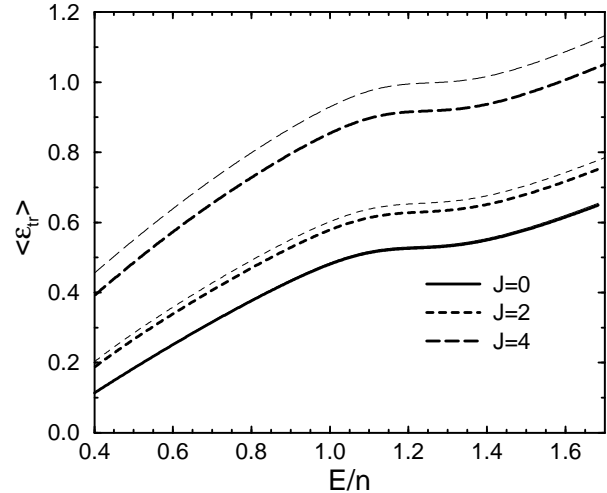}
\caption{Average total kinetic energy release in the dissociation of LJ$_{14}$
as a function of $E/n$ for 3 values of the initial angular momentum $J$, from
the predictions of PST using the simulated radial potential at $T=0.3$ (thick
lines) or the simpler $-C_6/r^6$ potential with $C_6=52$ (thin lines).}
\label{fig:etrav13_E}
\end{figure}

The variations of the average KER as a function of internal energy are displayed
in Fig.~\ref{fig:etrav13_E} for the same three values of angular momentum,
$J=0$, $J=2$, and $J=4$ LJ units. The results of the PST calculations are
represented in the energy range where an inflection occurs due to the melting
phase change in LJ$_{13}$.\cite{wa} The excitation energy where this inflection
takes place show a weak dependence over the radial potential used, as well as
a nearly constant value with increasing $J$. As angular momentum increases, the
average KER also increases, following the expected scaling law
$\langle \etr\rangle (J>0) \approx \langle\etr\rangle(J=0) + aJ^2$.
Looking now at the differences between the PST calculations, we notice that the use
of the $-C/r^6$ radial potential overestimates the average energy released, more and
more as $J$ increases. This quantitative difference can be explained from the
differences in the RDOS, as seen in Fig.~\ref{fig:gamma13}. For excitation energies
$E/n\sim 1$, $\langle\etr\rangle$ is less than about 1. In this range, 
the energy shift $\etr^{\rm min}$ plays a crucial role, and its overestimation
with the $r_0=0$ approximation is consistent with the larger average kinetic
energy released in Fig.~\ref{fig:etrav13_E}.

\begin{figure}[htb]
\setlength{\epsfxsize}{9cm}
\epsffile{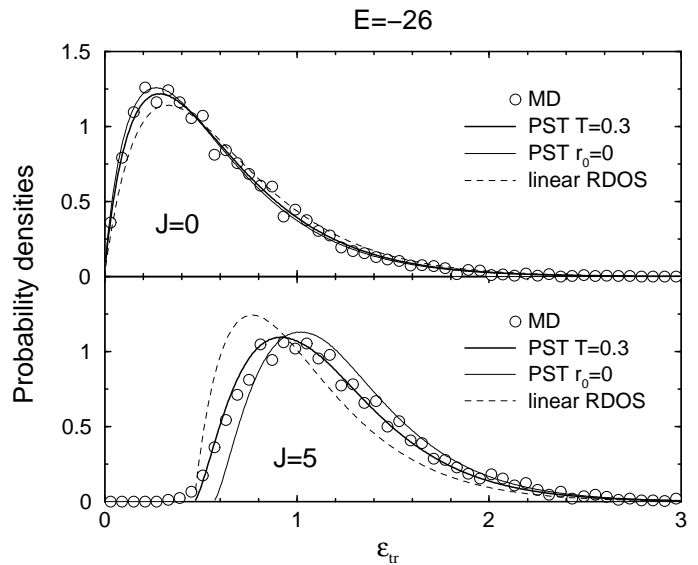}
\caption{Probability distributions of the total kinetic energy release
$\etr$ at $E=-26$ for two values of
$J$, in the dissociation of LJ$_{14}$: (a) $J$=0; (b) $J$=5. The dashed lines
refer to the linear approximation $\Gamma(\etr,J)\propto \etr-\etr^{\rm min}(J)$.}
\label{fig:dist13}
\end{figure}

Because Fig.~\ref{fig:etrav13_J} does not confidently discriminates between
the various approximations used in our application of PST, we turn to
the probability distribution of $\etr$ at given
total energy and angular momentum of the parent cluster, which carries more
precise information. In
Fig.~\ref{fig:dist13} we have represented these distributions obtained from
MD simulations and from PST, with the two radial potentials either with
$r_0=0$ and $C_6=4n$ or with the potential extracted from simulations close to
the melting point. The results are given for $J=0$ and $J=5$ LJ units. As can
be observed on the upper panel of this figure, the agreement between MD and PST
is excellent at $J=0$, and the two PST calculations give very similar data.
The rotational density of states roughly shows linear variations upon increasing
$\etr$ in Fig.~\ref{fig:gamma13}. We also show in Fig.~\ref{fig:etrav13_J} the
probability densities computed using the explicit linear approximation $\Gamma(\etr,J)
=\alpha(\etr - \etr^{\rm min})$.
Actually the slope constant $\alpha$ does not play any role in the KER distribution. 
This linear Weisskopf-like behavior\cite{weissk} leads to slight shift of the
distribution toward higher energies.

A nonzero angular momentum appears as a more stringent test for the statistical
theory, as we see on the lower panel of Fig.~\ref{fig:dist13} that the
agreement between MD and PST is significantly better with the calculation
performed using the radial potential corresponding to $T=0.3$. This is
consistent with the better agreement previously observed in
Fig.~\ref{fig:eav13}. In particular, the threshold value $\etr^{\rm min}$ of
$\etr$ at which the probability suddenly rises (near $\etr\sim 0.5$ LJ unit)
is well reproduced by PST at $T=0.3$, but is not in the approximation
$r_0=0$. Evaporations from rotating clusters are characterized by a nonzero
value of this threshold, which is due to the extra excitation energy needed
for the system to overcome the larger centrifugal barrier.
The influence of the shape of the rotational density of states on the
probability distribution of $\etr$ is also investigated using the linear
approximation. The difference with previous PST calculations is more
significant, which indicates the strong influence of the shape of the RDOS
in the vicinity of $\etr^{\rm min}$.

\begin{figure}[htb]
\setlength{\epsfxsize}{9cm}
\epsffile{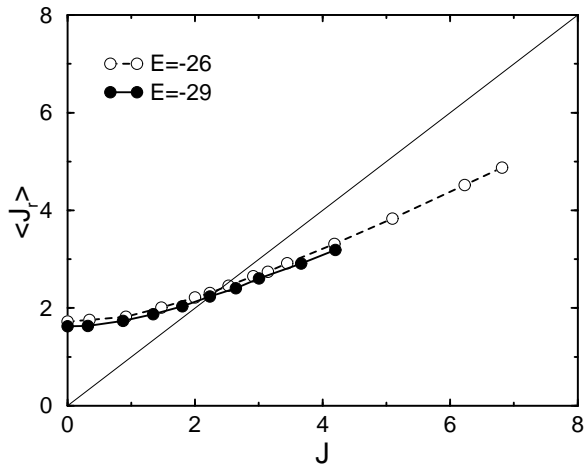}
\caption{Average angular momentum of the product subcluster as a function of
the initial angular momentum $J$ of the parent LJ$_{14}$, from MD
simulations at two total energies.}
\label{fig:jav13}
\end{figure}

Finally, we have represented in Fig.~\ref{fig:jav13} the average final angular
momentum $J_r$ of the product cluster as a function of the initial $J$, at
two total internal energies, from molecular dynamics simulations. 
On this figure the line $J_r-J=0$ is also drawn. Depending on the initial
angular momentum, the product cluster can gain or lose some of its rotational
velocity. These rotational cooling and rotational heating effects occur for
$J\lesssim J_0=2.3$ LJ units and $J\gtrsim J_0$, respectively. The threshold
value $J_0$ weakly depends on the total internal energy, in the rather limited
range investigated here.\cite{stace91} This can be understood using the
following simple arguments. At low $J$, evaporating an atom induces an orbital
momentum $\vec L$, which is nearly balanced by the angular momentum
$\vec J_r$. Hence angular momentum increases for initially small values of $J$.
On the other hand, rapidly rotating clusters lose a part of their angular
velocity by emitting one atom, and $J$ tends to decrease upon evaporation.

\subsection{\boldmath LJ$_8\longrightarrow$LJ$_7+$LJ\unboldmath}

\begin{figure}[htb]
\setlength{\epsfxsize}{9cm}
\epsffile{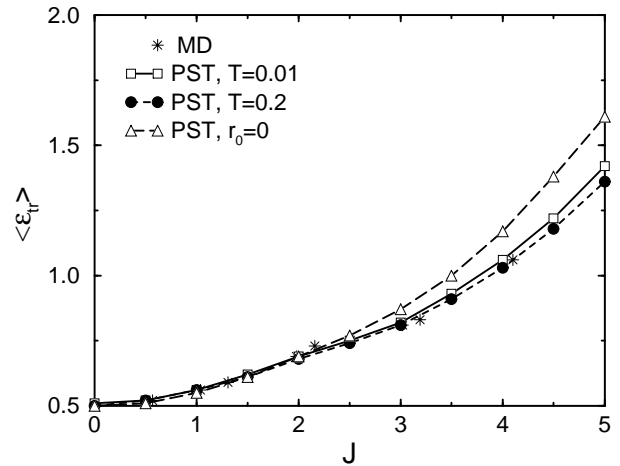}
\caption{Average total kinetic energy release in the dissociation of LJ$_8$
as a function of angular momentum $J$ at $E$=-10.21. Comparison between MD simulation and
phase space theory using different radial potentials.}
\label{fig:etrav7_J}
\end{figure}

We now turn to the evaporation statistics in the smaller LJ$_8$ cluster. The
product cluster LJ$_7$ is nonspherical not only in its lowest energy structure,
but also in any of his three other stable isomers. The average kinetic energy
released during dissociation is represented in Fig.~\ref{fig:etrav7_J} versus
the total angular momentum of the parent cluster $J$. The values plotted are
the results of MD simulations as well as the predictions of PST under various
approximations concerning the radial potential. As in the previous paragraph,
we have considered the simple $-C_6/r^6$ case with $C_6=4n$, and two more
realistic potentials extracted from Monte Carlo simulations at $T=0.01$ and
$T=0.2$, respectively. The latter value is close to the isomerization point in
this system. The effective rotational constant was taken as the average over
the different instantaneous values at the corresponding temperatures. They are
also given in Table~\ref{tab:c6r0}. As in Fig.~\ref{fig:etrav13_J}, the general
agreement between MD and PST is good, but we notice that the discrepancy is
larger for the PST calculation with $r_0=0$. The effect of temperature on the
radial potential is weak in this case. This may be partly due to the lower melting point
of this system, but should be correlated with the similar radial potentials felt
by the leaving atom, as represented in Fig.~\ref{fig:vradial}.

\begin{figure}[htb]
\setlength{\epsfxsize}{9cm}
\epsffile{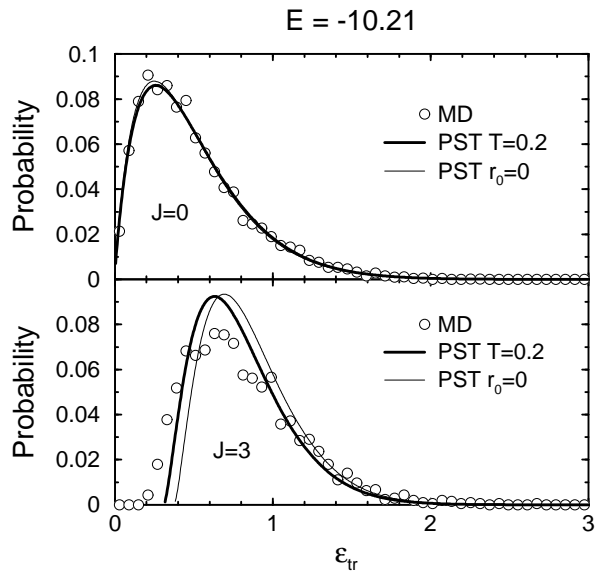}
\caption{Probability distributions of the total kinetic energy release
$\etr$ at $E=-10.21$ for two values of
$J$, in the dissociation of LJ$_8$: (a) $J$=0; (b) $J$=3.}
\label{fig:dist7}
\end{figure}

The probability distribution of kinetic energy released is reported in
Fig.~\ref{fig:dist7} at the total energy $E=-10.21$ LJ units, and at zero or
nonzero angular momentum of the parent cluster. As was the case for the bigger
cluster, PST reproduces very accurately the results of MD simulations at $J=0$,
and the two calculations with the different radial potentials yield essentially
similar data. In contrast, the distributions at $J=3$ differs somewhat from the
simulation results. In particular, the general shape predicted by PST is too
sharp with respect to MD, and the value of $\etr$ where the probability starts
to increase is too high by about 50\%. This error further increases when using
the alternative radial potential corresponding to $r_0=0$. The discrepancies
observed here should be mainly due to the erroneous assumption of a spherical
product.

Nevertheless, the global behaviors of both the KER and its probability
distribution remain correctly reproduced by phase space theory, indicating that
this statistical approach captures all the important physical and chemical
ingredients of unimolecular dissociation in weakly bound systems, especially
the conservation of angular momentum. 

\section{Conclusion}
\label{sec:ccl}

Unimolecular decay in large atomic systems is easily treated using simple
theories such as the RRKM or Weisskopf-Engelking statistical approaches.
Phase space theory not only includes possible anharmonic effects, but also the
rigorous constraints on angular momentum, through general expressions
for the differential rates of dissociation. Weerasinghe and Amar\cite{wa} were
the first to show clearly that PST is qualitatively and quantitatively accurate
in predicting rate constants and energetic distributions in the evaporation of
nonrotating atomic clusters. Building upon their seminal paper, we have
extended their work to the more general case of a finite angular momentum in
the parent cluster. For this we calculated exactly the
rotational density of states
in the case of an interaction between the product cluster and the dissociating
atom having the form $-C/r^p$. The implementation to other forms has also been
given. The anharmonic vibrational densities of states were calculated using
Monte Carlo simulations on the effective rovibrational energy
surface,\cite{mcrot} and the
radial potential was calculated using an extension of the recent Wang-Landau
algorithm.\cite{wl,fcwl}

We have tested the applicability of PST to the case of unimolecular evaporation
in the LJ$_{14}$ and LJ$_8$ clusters. These two different cases allow us to
question the hypothesis of a sphere$+$atom collision underlying the statistical
formalism. We have shown that PST was quantitatively accurate in predicting the
distribution and average value of the kinetic energy released during
dissociation, especially after considering the radial potential calculated at
temperatures close to the melting point, where dissociation actually occurs on
the time scale of MD. Taking the simple form $-C/r^6$ introduces some extra
errors, in particular at large energies and angular momenta. We have also seen
that dissociation in LJ$_8$ was less well described using PST in the sphere$+
$atom assumption. Beyond this approximation, one could generalize the present
formalism to the case of ellipsoid or even triaxial shapes. Extension to
molecular systems is also possible, provided that the internal degrees of
freedom of the dissociating molecule are correctly accounted for.

To bridge the gap between the results obtained in the present work and the
experimental concerns, one must extract the separate translational and
rotational contributions to the total kinetic energy released during
evaporation. This separation was achieved previously in the case of
nonrotating parent clusters.\cite{pbepjd} As a first step, it would be useful to
extend the present effort to the characterization of the angular momenta distribution
after dissociation of an initially rotating system. 
The starting distribution of $J$ could there be either a delta function (as in
the present work) or a thermalized Boltzmann distribution $P(J)\propto J^2
\exp(-BJ^2/k_BT)$. Work along these lines is presently in progress.

\section*{Acknowledgments}

The authors wish to thank the GDR {\it{Agr\'egats, Dynamique et R\'eactivit\'e}} for financial support.

\end{document}